\documentclass[12pt]{iopart}

\usepackage{amssymb}
\usepackage{amsthm}

 \newtheorem{theorem}{Theorem}[section]
        \newtheorem{lemma}[theorem]{Lemma}

     \newtheorem{definition}[theorem]{Definition}
        
\newcommand{\ud}{\mathrm{d}}
\begin{document}

\title[]{On Measure Theoretic definitions of Generalized Information
  Measures and Maximum Entropy Prescriptions}

\author{Ambedkar Dukkipati, M Narasimha Murty\footnote{Corresponding author} and
Shalabh Bhatnagar}

\address{Department of Computer Science and Automation,
Indian Institute of Science, Bangalore-560012, India.}
\ead{\mailto{ambedkar@csa.iisc.ernet.in},
\mailto{mnm@csa.iisc.ernet.in}, \mailto{shalabh@csa.iisc.ernet.in}}

%----------------------------------------
\begin{abstract}
 	Though Shannon entropy of a probability measure $P$, defined
        as $- \int_{X} \frac{\ud P}{\ud \mu} \ln \frac{\ud P}{\ud
        \mu} \, \ud \mu$ on a measure space $(X, \mathfrak{M},\mu)$, does not
        qualify itself as an information measure (it is not a natural
        extension of the discrete case), maximum entropy (ME)
        prescriptions in the measure-theoretic case are consistent with that of
        discrete case.  
        In this paper, we study the
        measure-theoretic definitions of generalized information
        measures and discuss the ME prescriptions. We present two
        results in this regard: (i) we prove that, as in
        the case of classical relative-entropy, the measure-theoretic
        definitions of generalized relative-entropies, R\'{e}nyi and
        Tsallis, are natural extensions of their respective discrete
        cases, (ii) we show that, ME prescriptions of
        measure-theoretic Tsallis entropy are consistent with the
        discrete case.
\end{abstract}

%Uncomment for PACS numbers title message
% Keywords required only for MST, PB, PMB, PM, JOA, JOB? 
%\vspace{2pc}
%\noindent{\it Keywords}: Article preparation, IOP journals
% Uncomment for Submitted to journal title message
%\submitto{\JPA}
% Comment out if separate title page not required
\maketitle

%=========================Introduction===========================
\section{Introduction}
\label{Section:Introduction}
        Shannon measure of information was developed
        essentially for the case when the random variable takes a
        finite number of values. However, in the literature, one often
        encounters an extension of Shannon entropy in the discrete 
        case to the case
        of a one-dimensional random variable with density function $p$ 
        in the form~(e.g \cite{ShannonWeawer:1949:TheMathematicalTheoryOfCommunication,Ash:1965:InformationTheory})  
        \begin{displaymath}
          S(p) = - \int_{- \infty}^{+ \infty} p(x) \ln p(x)\, \ud x \enspace.
        \end{displaymath}
        This entropy in the continuous case 
        as a pure-mathematical formula (assuming convergence of
        the integral and absolute continuity of the density $p$ with
        respect to Lebesgue measure) resembles Shannon entropy in the
        discrete case, but can not be used as a measure of
        information. First, it is not a natural extension of Shannon
        entropy in the discrete case, since it is not the limit of the sequence
        finite discrete entropies corresponding to pmf which
        approximate the pdf $p$. Second, it is not strictly positive.

        Inspite of these short comings, one can still use the
        continuous entropy functional in conjunction with the principle of maximum
        entropy where one wants to find a probability density function
        that has greater uncertainty than any other distribution
        satisfying a set of given constraints. Thus, in this use of
        continuous measure one is interested in it as a measure of
        relative uncertainty, and not of absolute uncertainty. This
        is where one can relate maximization of Shannon entropy to the
        minimization of Kullback-Leibler relative-entropy
        (see~\cite[pp. 55]{KapurKesavan:1997:EntropyOptimizationPrinciples}).
%        It
%        is well known that the continuous version of
%        KL-entropy defined for two probability density functions $p$
%        and $r$ as,
%        \begin{displaymath}
%         I(p\|r) = \int_{- \infty}^{+ \infty} p(x) \ln
%        \frac{p(x)}{r(x)} \, \ud x \enspace,
%        \end{displaymath}   
%        is indeed a natural generalization of same in the discrete
%        case.

        Indeed, during the early stages of development of
        information theory, the important paper 
        by Gelfand, Kolmogorov and Yaglom~\cite{GelfandKolmogorovYaglom:1956:OnTheGeneralDefinitionOfTheAmountOfInformation} 
        called attention to the case of defining entropy functional on
        an arbitrary measure space $(X, \mathfrak{M},\mu)$. 
	In this respect, Shannon entropy of a probability density function $p:X 
        \rightarrow {\mathbb{R}}^{+}$ can be written as,
        \begin{displaymath}
          S(p) = - \int_{X} p(x) \ln p(x) \, \ud \mu \enspace.
        \end{displaymath}  
        One can see from the above definition that the concept of
        ``the entropy of a pdf'' is a misnomer: there
        is always another measure $\mu$ in  the background. In the
        discrete case considered by Shannon, $\mu$ is the cardinality
        measure\footnote{Counting or cardinality measure $\mu$ on a
          measurable space $(X,\mathfrak{M})$, when is $X$ is a
          finite set and $\mathfrak{M} = 2^{X}$, is defined as $\mu(E)
          = \# E$, $\forall E \in \mathfrak{M}$.}~\cite[pp. 19]{ShannonWeawer:1949:TheMathematicalTheoryOfCommunication};
        in the continuous case considered by both Shannon and Wiener,
        $\mu$ is the Lebesgue
        measure cf.~\cite[pp. 54]{ShannonWeawer:1949:TheMathematicalTheoryOfCommunication}
        and 
        \cite[pp. 61, 62]{Wiener:1948:Cybernetics}.
         All entropies are
        defined with respect to some measure
        $\mu$,
        as Shannon and Wiener both emphasized in~\cite[pp.57,
        58]{ShannonWeawer:1949:TheMathematicalTheoryOfCommunication} 
        and~\cite[pp.61, 62]{Wiener:1948:Cybernetics} respectively.

        This case was studied independently
        by Kallianpur~\cite{Kallianpur:1960:OnTheAmountOfInformationContainedInASingmaField} 
        and Pinsker~\cite{Pinsker:1960:InformationAndInformationStability},
        and perhaps others were guided by the earlier work
        of Kullback~\cite{KullbackLeibler:1951:OnInformationAndSufficiency},
        where one would define entropy in terms of Kullback-Leibler
        relative entropy. Unlike Shannon entropy, measure-theoretic
        definition of KL-entropy is a natural extension of definition
        in the discrete case. 

	%In this respect,
        %the Gelfand-Yaglom-Perez  theorem
        %(GYP-theorem)~\cite{GelfandYaglom:1959:CalculationOfTheAmountOfInformation_Etc,Perez:1959:InformationTheoryWithAbstractAlphabets,Dobrushin:1959:GeneralFormulationsOfShannonsbasicTheorems}
        %plays an important role, which equips measure-theoretic
        %KL-entropy with a fundamental definition. The main
        %contribution of this chapter is to prove GYP-theorem for 
	%R\'{e}nyi relative-entropy of order $\alpha >1$, which can be
        %extended to Tsallis relative-entropy.

	%Before proving GYP-theorem for R\'{e}nyi relative-entropy,
	In this paper we present the measure-theoretic definitions of
	generalized information measures and show that as in
        the case of KL-entropy, the measure-theoretic
        definitions of generalized relative-entropies, R\'{e}nyi and
        Tsallis, are natural extensions of their respective discrete
        cases. We discuss the ME prescriptions for generalized
	entropies and show that ME prescriptions of
        measure-theoretic Tsallis entropy are consistent with the
        discrete case, which is true for measure-theoretic
	Shannon-entropy. 

	Rigorous studies of the Shannon and KL entropy functionals in
	measure spaces can be found in the papers by
        Ochs~\cite{Ochs:1976:BasicPropertiesOfTheGeneralizedBoltzmann-Gibbs-ShannonEntropy}
        and by
        Masani~\cite{Masani:1992:TheMeasureTheoreticAspectsOfEntropy_Part_1,Masani:1992:TheMeasureTheoreticAspectsOfEntropy_Part_2}.
        Basic measure-theoretic aspects of classical information measures can be
        found
        in~\cite{Pinsker:1960:InformationAndInformationStability,Guiasu:1977:InformationTheoryWithApplications,Gray:1990:EntropyAndInformationTheory}.
%        in~\cite[Chapter~2]{Guiasu:1977:InformationTheoryWithApplications}
%        and~\cite[Chapter~5]{Gray:1990:EntropyAndInformationTheory}.

        We review the measure-theoretic formalisms for classical
        information measures in 
        \S~\ref{Section:ME:MeasureTheoreticDefinitionsOfInformationMeasures}
        and extend these definitions to generalized
        information measures in
        \S~\ref{Section:ME:MeasureTheoreticDefinitionsOfGeneralizedInformationMeasures}. In
        \S~\ref{Section:ME:MaximumEntropyAndCanonicalDistributions} we
        present the ME prescription for Shannon entropy followed by
        prescriptions for 
        Tsallis entropy in
        \S~\ref{Section:ME:ME-prescriptionForTsallisEntropy}. We
        revisit measure-theoretic definitions of generalized entropic
        functionals in
        \S~\ref{Section:ME:MeasureTheoreticDefinitions_Revisited} and
        present some results.

%================================Section:==========================================
\section{Measure-Theoretic definitions of Classical Information Measures}
\label{Section:ME:MeasureTheoreticDefinitionsOfInformationMeasures}
%	Information measures like entropy, mutual information,
%	conditional entropy, and conditional mutual information
%	etc., can be expressed in terms of KL-entropy and hence
%	the measure-theoretic analogs of these measures will follow
%	from the measure-theoretic definition of KL-entropy.
%	In this section, we study the measure-theoretic
%	definitions of KL-entropy and its relation to entropy in this
%	case.  
  %-----------------------------SubSection-------------------------------------
  \subsection{Discrete to Continuous}
   \label{SubSection:ME:DiscreteToContinuous}
        \noindent
        Let $p:[a,b] \rightarrow {\mathbb{R}}^{+}$ be a probability
        density function,  where $[a,b] \subset \mathbb{R}$. That is,
        $p$ satisfies
        \begin{displaymath}
        p(x) \geq 0, \:\:\: \forall x \in [a,b] \:\:\: \mathrm{and}\:\:\:
        \int_{a}^{b} p(x) \, \ud x =1 \enspace.
        \end{displaymath}
        In trying to define entropy in the continuous case, the
        expression of Shannon entropy was automatically extended by
        replacing the sum in the 
	Shannon entropy discrete case by the
        corresponding integral. We obtain, in this way, Boltzmann's
        H-function (also known
  	as differential entropy in information theory),
%	~\cite{Grad:1965:OnBoltzmannsH-Theorem}: reference for
%        Boltzmann-H function
        \begin{equation}
        \label{Equation:ME:ContinuousEntropy}
        S(p) = - \int_{a}^{b} p(x) \ln p(x) \, \ud x \enspace.
        \end{equation}
        But the ``continuous entropy'' given
        by~(\ref{Equation:ME:ContinuousEntropy}) is not a natural
        extension of definition in discrete case in the sense that, it
        is not the limit of
        the finite discrete entropies corresponding to a sequence of
        finer partitions of the interval $[a,b]$ whose norms tend to
        zero. We can show this by a counter example. 
        Consider a uniform probability distribution 
        on the interval $[a,b]$, having the probability density
        function
        \begin{displaymath}
        p(x) = \frac{1}{b-a}\enspace, \:\:\:\:\: x \in [a,b] \enspace.
        \end{displaymath}
        The continuous
        entropy~(\ref{Equation:ME:ContinuousEntropy}), in this case will be
        \begin{displaymath}
        S(p) = \ln (b - a) \enspace. 
        \end{displaymath}
        On the other hand, let us consider a finite partition of the the interval
        $[a,b]$ which is composed of $n$ equal subintervals, and let
        us attach to this partition the finite discrete uniform
        probability distribution whose corresponding entropy will be,
        of course,
        \begin{displaymath}
        S_{n}(p) = \ln n \enspace.
        \end{displaymath}
        Obviously, if $n$ tends to infinity, the discrete entropy
        $S_{n}(p)$ will tend to infinity too, and not to $\ln (b-a)$;
        therefore $S(p)$ is not the limit of $S_{n}(p)$, when $n$ tends
        to infinity. Further, one can observe that $\ln (b-a)$ is negative  
        when~$b-a <1$.

	Thus, strictly speaking
        continuous entropy~(\ref{Equation:ME:ContinuousEntropy}) cannot 
        represent a measure of uncertainty since uncertainty should
        in general be positive.
	We are able to prove the ``nice'' properties only for the
        discrete entropy, therefore, it
        qualifies as a ``good'' measure of information (or
        uncertainty) supplied by an random experiment. The ``continuous
        entropy'' not being the limit of the discrete
        entropies, we cannot extend the so called nice properties to
        it.

        Also, in physical applications, the coordinate $x$ in
        (\ref{Equation:ME:ContinuousEntropy}) represents an abscissa,
        a distance from a fixed reference point. This distance $x$ has
        the dimensions of length. Now, with the density function
        $p(x)$, one can specify the probabilities of an event $[c,d)
        \subset [a,b]$ as $\int_{c}^{d} p(x) \, \ud x$, one has to
        assign the dimensions ${(\mbox{length})}^{-1}$, since
        probabilities are dimensionless. Now for $0 \leq z < 1$, one
        has the series expansion
        \begin{equation}
          - \ln (1-z) = z + \frac{1}{2}z^{2} + \frac{1}{3}z^{3}+
          \ldots \enspace,
        \end{equation}
        it is necessary that the argument of the logarithm function
        in~(\ref{Equation:ME:ContinuousEntropy}) be
        dimensionless.
	Hence the formula (\ref{Equation:ME:ContinuousEntropy}) is
        then seen to be dimensionally incorrect, since the argument of
        the logarithm on its right hand side has the dimensions of a
        probability
        density~\cite{Smith:2001:SomeObservationsOnTheConceptsOfInformationTheoreticEntropy}.
        Although
        Shannon~\cite{Shannon:1948:MathematicalTheoryOfCommunication_BellLabs} 
        used the formula (\ref{Equation:ME:ContinuousEntropy}), he
        does note its lack of invariance with respect to changes in
        the coordinate system.

        In the context of maximum entropy principle
        Jaynes~\cite{Jaynes:1968:PriorProbabilities} 
        addressed this problem and suggested the formula,
        \begin{equation}
        \label{Equation:ME:JaynesSuggestion}  
          S'(p) = - \int_{a}^{b} p(x) \ln \frac{p(x)}{m(x)}\, \ud x \enspace,
        \end{equation}
        in the place of (\ref{Equation:ME:ContinuousEntropy}),  
        where $m(x)$ is a prior function. Note that when $m(x)$ is probability density
        function, (\ref{Equation:ME:JaynesSuggestion}) is nothing but
        the relative-entropy. However, if we choose $m(x) = c$, a constant
        (e.g \cite{ZellnerHighfield:1988:CalculationOfMaximumEntropyDistributions}),
        we get 
        \begin{displaymath}
          S'(p) = S(p) - \ln c \enspace,
        \end{displaymath}
        where $S(p)$ refers to the continuous
        entropy (\ref{Equation:ME:ContinuousEntropy}).
        Thus, maximization of $S'(p)$ is equivalent to maximization of
        $S(p)$.
	Further discussion on estimation of probability
        density functions by ME-principle in the continuous case can be found in  
        \cite{LazoRathie:1978:OnTheEntropyOfContinuousProbabilityDistributions,ZellnerHighfield:1988:CalculationOfMaximumEntropyDistributions,Ryu:1993:MaximumEntropyEstimationOfDensityAndRegressionFunction}.

        Prior to that, Kullback~\cite{KullbackLeibler:1951:OnInformationAndSufficiency} too
        suggested that in the measure-theoretic definition of entropy,
        instead of examining the entropy 
        corresponding to only on given measure, we have to compare the
        entropy inside a whole class of measures.

  %-----------------------SubSection------------------------------------
  \subsection{Classical information measures}
  \label{SubSection:ME:ClassicalInformationMeasures}

        \noindent
        Let $(X,\mathfrak{M},\mu)$ be a measure space. $\mu$
        need not be a probability measure unless otherwise specified.
        Symbols $P$, $R$ will denote probability measures on
        measurable space $(X,\mathfrak{M})$ and $p$, $r$  
        denote $\mathfrak{M}$-measurable functions on $X$.
        An $\mathfrak{M}$-measurable function $p:X \rightarrow
        {\mathbb{R}}^{+}$ is said to be a probability 
        density function (pdf) if $\int_{X} p \, \ud \mu = 1$.

        In this general setting, Shannon entropy $S(p)$ of pdf $p$ is
        defined as follows~\cite{Athreya:1994:EntropyMaximization}. 
        %DEFINITION: Shannon entropy for pdf
        \begin{definition}
        \label{Definition:ME:ShannonEntropy_Measuretheroetic_pdf}
        Let $(X,\mathfrak{M},\mu)$ be a measure space and 
        $\mathfrak{M}$-measurable function $p:X \rightarrow  
        {\mathbb{R}}^{+}$ be pdf. Shannon entropy of $p$
        is defined as
        \begin{equation}
         \label{Equation:ME:ShannonEntropyOf-pdf} 
        S(p) = - \int_{X} p \ln p \, \ud \mu \enspace,
        \end{equation}
        provided the integral on right exists.
        \end{definition}%EndDefinition
        Entropy functional $S(p)$ defined in (\ref{Equation:ME:ShannonEntropyOf-pdf}) can be
        referred to as entropy of the probability measure 
        $P$, in the sense that the measure $P$ is induced by $p$,
        i.e.,
        \begin{equation}
        \label{Equation:ME:ProbabilityMeasureInducedByaPdf}  
          P(E) = \int_{E} p(x) \, \ud \mu(x) \enspace, \:\:\:\:\:
          \forall E \in \mathfrak{M} \enspace.
        \end{equation}
	This reference is consistent\footnote{Say
        $p$ and 
        $r$ are two pdfs and $P$ and $R$ are corresponding
        induced measures on measurable space $(X,\mathfrak{M})$ such
        that $P$ and $R$ are identical, i.e., $\int_{E} p \,
        \ud \mu = \int_{E} r \, \ud \mu$, $\forall E \in \mathfrak{M}$. Then
        we have $p \stackrel{\mathrm{a.e}}{=} r$ and hence
        $ -\int_{X} p \ln p \, \ud \mu = -\int_{X} r \ln r \, \ud
        \mu$.} because the probability measure
        $P$ can be identified {\it a.e} by the pdf $p$.

        Further, the definition of the probability measure $P$
        in (\ref{Equation:ME:ProbabilityMeasureInducedByaPdf}), allows us
        to write entropy functional
        (\ref{Equation:ME:ShannonEntropyOf-pdf}) 
        as, 
        \begin{equation}
        \label{Equation:ME:ShannonEntropyOf-PM-inducedBy-pdf}
        S(p) = - \int_{X} \frac{\ud P}{\ud \mu} \ln \frac{\ud P}{\ud
        \mu} \, \ud \mu \enspace,
        \end{equation}
        since (\ref{Equation:ME:ProbabilityMeasureInducedByaPdf})
        implies\footnote{If a 
        nonnegative measurable function $f$ induces a measure $\nu$ on
        measurable space $(X,\mathfrak{M})$ with respect to a measure
        $\mu$, defined as $\nu(E) = \int_{E} f \, \ud \mu, \:\:\: \forall E \in
        \mathfrak{M}$ then $\nu \ll \mu$. Converse is given by
        Radon-Nikodym theorem~\cite[pp.36, Theorem
          1.40(b)]{Kantorovitz:2003:IntroductionToModernAnalysis}.} $P
        \ll \mu$, and pdf $p$ is the
        Radon-Nikodym derivative of $P$ w.r.t $\mu$. 

        Now we proceed to the definition of Kullback-Leibler
        relative-entropy or KL-entropy for probability measures.
        %Definition:Kullback-Leibler Relative-Entropy1
        \begin{definition}
        \label{Definition:ME:RelativeEntropy_1}
        Let $(X,\mathfrak{M})$ be a measurable space. Let $P$ and $R$
        be two probability measures on $(X,\mathfrak{M})$. Kullback-Leibler
        relative-entropy  KL-entropy of $P$ relative to $R$ is
        defined as
        \begin{equation}
        \label{Equation:ME:RelativeEntropyOfProbabilityMeasures}
        I(P\|R) = \left\{ \begin{array}{ll}
        \displaystyle{\int_{X} \ln \frac{\ud P}{\ud R} \, \ud P }     &
        \:\:\:\:\:\textrm{if}\:\:\:\:\:  P \ll R \enspace, \\ \\
          +\infty   & \:\:\:\:\:\textrm{otherwise.}
           \end{array} \right.
        \end{equation}
        \end{definition}%EndDefinition:Kullback-Leiber Relative-Entropy1
	The divergence inequality
        $I(P\|R) \geq 0$ and $I(P\|R) =0$ if and only if $P=R$ can be
        shown in this case too.
        KL-entropy~(\ref{Equation:ME:RelativeEntropyOfProbabilityMeasures})
        also can be written as 
        \begin{equation}
        \label{Equation:ME:AnotherFormForRelativeEntropyOfProbabilityMeasures}  
        I(P\|R) = \int_{X} \frac{\ud P}{\ud R} \ln \frac{\ud P}{\ud R}
        \, \ud R \enspace.
        \end{equation}
        
        Let the $\sigma$-finite measure $\mu$ on $(X,\mathfrak{M})$
        such that $P \ll R \ll \mu$. Since $\mu$ is $\sigma$-finite, from
        Radon-Nikodym theorem, there exists a non-negative 
        $\mathfrak{M}$-measurable functions $p: X \rightarrow
        \mathbb{R}^{+}$ and $r: X \rightarrow \mathbb{R}^{+}$ unique
        $\mu$-{\em a.e}, such that
        \begin{equation}
	\label{Equation:ME:DefinitionOfPdf_p}
        P(E) = \int_{E} p \, \ud \mu \enspace, \:\:\: \forall E \in \mathfrak{M} \enspace,
        \end{equation}
        and
        \begin{equation}
	\label{Equation:ME:DefinitionOfPdf_r}
        R(E) = \int_{E} r \, \ud \mu \enspace, \:\:\: \forall E \in
        \mathfrak{M} \enspace.
        \end{equation}
        The pdfs $p$ and $r$ in (\ref{Equation:ME:DefinitionOfPdf_p})
        and (\ref{Equation:ME:DefinitionOfPdf_r}) (they are indeed 
        pdfs) are Radon-Nikodym 
        derivatives of probability measures $P$ and $R$ with respect
        to $\mu$, respectively, i.e., $p =\frac{\ud P}{\ud \mu}$ and 
        $r=\frac{\ud R}{\ud \mu}$.
        Now one can define relative-entropy of pdf $p$ w.r.t $r$ as
        follows\footnote{This follows from the chain rule for
        Radon-Nikodym derivative:
         \begin{displaymath}
           \frac{\ud P}{\ud R} \stackrel{\mathrm{a.e}}{=} \frac{\ud
             P}{\ud \mu} {\left( \frac{\ud R}{\ud \mu} \right)}^{-1}\enspace.
         \end{displaymath}  
        }.
        
       %Definition:KullbackLeibler Relative-Entropy2
        \begin{definition}
        \label{Definition:ME:RelativeEntropy_of_pdf}
        Let $(X,\mathfrak{M},\mu)$ be a measure space. Let
       $\mathfrak{M}$-measurable functions $p,r:X \rightarrow 
        {\mathbb{R}}^{+}$ be two pdfs. The KL-entropy of $p$
       relative to $r$ 
        is defined as
        \begin{equation}
         \label{Equation:ME:RelativeEntropy_of_pdf} 
        I(p\|r) = \int_{X} p(x) \ln \frac{p(x)}{r(x)} \, \ud \mu(x) \enspace,
        \end{equation}
        provided the integral on right exists.
        \end{definition}%EndDefinition:KullbackLeibler Relative-Entropy2

        As we have mentioned earlier, KL-entropy
        (\ref{Equation:ME:RelativeEntropy_of_pdf}) exist if the two 
        densities are absolutely continuous with respect to one
        another. On the real line the same definition can be written
        as
        \begin{displaymath}
        I(p\|r) = \int_{\mathbb{R}} p(x) \ln \frac{p(x)}{r(x)} \, \ud x \enspace,
        \end{displaymath}
        which exist if the densities $p(x)$ and $r(x)$ share the same support.
        Here, in the sequel we use the convention
        \begin{equation}
        \ln 0 = - \infty, \:\:\:\:\:\:\:\:\:\:\: \ln \frac{a}{0} = + \infty\:\:
        \mathrm{for any}\:\: a \in \mathbb{R}, \:\:\:\:\:\:\:\:\:\:\:
        0.(\pm \infty) = 0.
        \end{equation}
        
        Now we turn to the definition of entropy functional on a
        measure space.
        Entropy functional in 
        ~(\ref{Equation:ME:ShannonEntropyOf-PM-inducedBy-pdf}) is defined
        for a probability measure
        that is induced by a pdf. By the Radon-Nikodym theorem, one can
        define Shannon entropy for any arbitrary $\mu$-continuous probability measure as follows.
        %Definition: Shannon entropy of Probability measure
        \begin{definition}
         \label{Definition:ME:ShannonEntropy_of_ProbabiliyMeasure} 
         Let $(X,\mathfrak{M},\mu)$ be a $\sigma$-finite measure
        space. Entropy of any $\mu$-continuous probability measure $P$
        ($P \ll \mu$) is defined as
        \begin{equation}
        \label{Equation:ME:ShannonEntropy_of_ProbabilityMeasure}  
        S(P) = - \int_{X} \ln \frac{\ud P}{\ud \mu} \, \ud P  \enspace.
        \end{equation}
        \end{definition}
        Properties of entropy of a probability measure in the
        Definition~\ref{Definition:ME:ShannonEntropy_of_ProbabiliyMeasure} are
        studied in detail by
        Ochs~\cite{Ochs:1976:BasicPropertiesOfTheGeneralizedBoltzmann-Gibbs-ShannonEntropy} 
        under the name generalized Boltzmann-Gibbs-Shannon
        Entropy. In the literature, one can find notation of the form
        $S(P|\mu)$ to represent the entropy functional in
        (\ref{Equation:ME:ShannonEntropy_of_ProbabilityMeasure}) viz., the
        entropy of a  
        probability measure, to stress the role of the measure
        $\mu$ (e.g~\cite{Ochs:1976:BasicPropertiesOfTheGeneralizedBoltzmann-Gibbs-ShannonEntropy,Athreya:1994:EntropyMaximization}). Since
        all the information measures we define are with 
        respect to the measure $\mu$ on $(X, \mathfrak{M})$, we omit
        $\mu$ in the entropy 
        functional notation.

        By assuming $\mu$ as a probability measure in the
        Definition~\ref{Definition:ME:ShannonEntropy_of_ProbabiliyMeasure},
        one can relate Shannon entropy with Kullback-Leibler entropy
        as,
        \begin{equation}
        \label{Equation:ME:RelationBetweenMeasureTheoreticEntropyAndKullback} 
        S(P) = - I(P\|\mu) \enspace.
        \end{equation}
	Note that when $\mu$ is not a probability measure, the
        divergence inequality $I(P\|\mu) \geq 0$ need not be
        satisfied.

	A note on the
        $\sigma$-finiteness of measure $\mu$. In the definition of
        entropy functional we assumed that $\mu$ is a $\sigma$-finite
        measure. This condition was used by
        Ochs~\cite{Ochs:1976:BasicPropertiesOfTheGeneralizedBoltzmann-Gibbs-ShannonEntropy}, 
        Csisz\'{a}r~\cite{Csiszar:1969:OnGeneralizedEntropy}
        and
        Rosenblatt-Roth~\cite{Rosenblatt-Roth:1964:TheConceptOfEntropyInProbabilityTheory} 
        to tailor the measure-theoretic definitions. For all practical
        purposes and for most applications, this assumption is
        satisfied. (See
        \cite{Ochs:1976:BasicPropertiesOfTheGeneralizedBoltzmann-Gibbs-ShannonEntropy}
        for a discussion on the physical interpretation of measurable space
        $(X,\mathfrak{M})$ with $\sigma$-finite measure $\mu$ for
        entropy measure of the
        form~(\ref{Equation:ME:ShannonEntropy_of_ProbabilityMeasure}),
        and of the relaxation $\sigma$-finiteness 
        condition.) By relaxing this condition, more universal
        definitions of entropy functionals are studied
        by Masani~\cite{Masani:1992:TheMeasureTheoreticAspectsOfEntropy_Part_1,Masani:1992:TheMeasureTheoreticAspectsOfEntropy_Part_2}.

%        In this thesis we will not go into those details.  

  %---------------------------------------------------
  \subsection{Interpretation of Discrete and Continuous Entropies in
  terms of KL-entropy}
  \label{SubSection:ME:MeasureTheoreticCasesinDiscrete}	
	\noindent
        First, let us consider discrete case of $(X, \mathfrak{M},
	\mu)$, where $X= \{x_{1}, \ldots, x_{n} \} $, $\mathfrak{M} =
	2^{X}$ and $\mu$ is a cardinality probability measure. Let $P$
	be any probability measure on $(X, \mathfrak{M})$. Then $\mu$
  	and $P$ can be specified as follows.
        \begin{displaymath}
        \mu \mbox{:} \:\:\: {\mu}_{k} = \mu(\{x_{k}\})  \geq 0, \:\:k = 1,
        \ldots, n, \:\:\:\sum_{k=1}^{n} 
        \mu_{k} =1 \enspace, \:\:\: %\mbox{and}
        \end{displaymath}
	and
        \begin{displaymath}
        P \mbox{:}\:\:\:  P_{k} = P(\{x_{k}\}) \geq 0 , \:\:k =1,
        \ldots, n, \:\:\: \sum_{k=1}^{n} P_{k} =1 \enspace.
        \end{displaymath}
        The probability measure $P$ is absolutely
        continuous with respect to the probability measure $\mu$ if
        $\mu_{k} =0$ implies $P_{k} =0$ for any $k=1,\ldots n$. The
        corresponding Radon-Nikodym 
        derivative of $P$ with respect to $\mu$ is given by
        \begin{displaymath}
                \frac{\ud P}{\ud \mu}(x_{k}) = \frac{P_{k}}{\mu_{k}}, \,
                k = 1, \ldots n \enspace.
        \end{displaymath}
        The measure-theoretic entropy $S(P)$
        (\ref{Equation:ME:ShannonEntropy_of_ProbabilityMeasure}),
        in this case, can be written as 
        \begin{displaymath}
        S(P) = - \sum_{k=1}^{n} P_{k}\ln \frac{P_{k}}{\mu_{k}} =
        \sum_{k=1}^{n} P_{k} \ln \mu_{k} - \sum_{k=1}^{n} P_{k} \ln
        P_{k} \enspace.
        \end{displaymath}
        If we take referential
        probability measure $\mu$ as a uniform probability
        distribution on the set $X$, i.e. $\mu_{k} = \frac{1}{n}$, we obtain
        \begin{equation}
        \label{Equation:ME:RelativionBetweenMeasureTheoreticAndDiscreteEntropies}
        S(P) = S_{n}(P) - \ln n \enspace,
        \end{equation}
	where $S_{n}(P)$ denotes the Shannon entropy of pmf $P =
        (P_{1}, \ldots, P_{n})$ and $S(P)$ denotes
	the 
        measure-theoretic entropy in the discrete case.

	Now, lets consider the continuous case of
	$(X,\mathfrak{M},\mu)$, where $X = [a,b] \subset \mathbb{R}$,
	$\mathfrak{M}$ is set of Lebesgue measurable sets of $[a,b]$,
	and $\mu$ is the Lebesgue probability measure. In this case
	$\mu$ and $P$ can be specified as follows.
        \begin{displaymath}
        \mu \mbox{:}\:\:\: \mu(x) \geq 0 , x \in
	[a,b], \ni \mu(E) = \int_{E} \mu(x) \, \ud x, \forall E \in
	\mathfrak{M}, \: \int_{a}^{b} \mu(x)\, \ud x  =1 \enspace, 
        \end{displaymath}
	and
        \begin{displaymath}
        P \mbox{:}\:\:\:  P(x) \geq 0 , x \in
	[a,b], \ni  P(E) = \int_{E} P(x) \, \ud x, \forall E \in \mathfrak{M}, \:\int_{a}^{b} P(x)\, \ud x =1 \enspace.
        \end{displaymath}
	Note the abuse of notation in the above specification of
	probability measures $\mu$ and $P$, where we have used the same
	symbols for both measures and pdfs.

        The probability measure $P$ is absolutely continuous with
        respect to the probability measure $\mu$, if $\mu(x)=0$ on a
        set of a positive Lebesgue measure implies
        that $P(x)=0$ on the same 
        set. The Radon-Nikodym derivative of the probability measure
        $P$ with respect to the probability measure $\mu$ will be 
        \begin{displaymath}
                \frac{\ud P}{\ud \mu}(x) = \frac{P(x)}{\mu(x)} \enspace.
        \end{displaymath}
        Then the measure-theoretic entropy $S(P)$ in this case
	can be written as 
        \begin{displaymath}
        S(P) = - \int_{a}^{b} P(x) \ln \frac{P(x)}{\mu(x)} \, \ud x
        \enspace. 
        \end{displaymath}
        If we take referential probability measure $\mu$ as a uniform
	distribution, i.e. $\mu(x) = \frac{1}{b-a}$, $x \in [a,b]$,
        then we obtain
        \begin{displaymath}
        \label{Equation:ME:RelativionBetweenMeasureTheoreticAndContinuousEntropies}
        S(P) = S_{[a,b]}(P) - \ln (b-a) \enspace,
        \end{displaymath}
	where $S_{[a,b]}(P)$ denotes the Shannon entropy of pdf
        $P(x)$, $x \in [a,b]$ (\ref{Equation:ME:ContinuousEntropy})
	and $S(P)$ denotes the measure-theoretic entropy in the
	continuous case. 

        Hence, one can conclude that
        measure theoretic entropy $S(P)$ defined for a probability measure $P$ on
        the measure space $(X,\mathcal{M},\mu)$, is equal to both Shannon 
        entropy in the discrete and continuous case case up to an
        additive constant, when the reference measure $\mu$ is chosen as a uniform
        probability distribution.
	On the other hand, one can see that measure-theoretic KL-entropy,
        in discrete and continuous cases are equal to its discrete and
        continuous definitions.
        
        Further, from
        (\ref{Equation:ME:RelationBetweenMeasureTheoreticEntropyAndKullback}) and
        (\ref{Equation:ME:RelativionBetweenMeasureTheoreticAndDiscreteEntropies}),
        we can write Shannon Entropy in terms Kullback-Leibler
        relative entropy
        \begin{equation}
        S_{n}(P) = \ln n - I(P \| \mu) \enspace.
        \end{equation}
        Thus, Shannon entropy appearers as being (up to an additive
        constant) the variation of information when we pass from the
        initial uniform probability distribution to new probability
        distribution given by $P_{k} \geq 0$, $\sum_{k=1}^{n} P_{k}
        =1$, as any such probability distribution is obviously
        absolutely continuous with respect to the uniform discrete
        probability distribution.
        Similarly, by
        (\ref{Equation:ME:RelationBetweenMeasureTheoreticEntropyAndKullback})
        and
        (\ref{Equation:ME:RelativionBetweenMeasureTheoreticAndContinuousEntropies})
        the relation between Shannon entropy and Relative entropy in
        discrete case 
        we can write Boltzmann H-function in terms of Relative entropy 
        as
        \begin{equation}
        S_{[a,b]}(p) = \ln (b-a) - I(P \| \mu) \enspace.
        \end{equation}
        Therefore, the continuous entropy or Boltzmann H-function
        $S(p)$ may be interpreted as being (up to an additive
        constant) the variation of information when we pass from the
        initial uniform probability distribution on the interval
        $[a,b]$ to the new probability measure defined by the
        probability distribution function $p(x)$ (any such 
        probability measure is absolutely continuous with respect to
        the uniform probability distribution on the interval
        $[a,b]$).

	Thus, KL-entropy equips one with unitary interpretation of both
	discrete entropy and continuous entropy.
        One can utilize Shannon entropy in the continuous case,
        as well as Shannon entropy in the discrete
        case, both being interpreted as the variation of information
        when we pass from the initial uniform distribution to the
        corresponding probability measure.

        Also,
        since measure theoretic entropy is equal to the discrete and
        continuous entropy upto an additive constant, ME prescriptions
        of measure-theoretic Shannon entropy are consistent with
        discrete case and the continuous case.

%=======================Section:================================
\section{Measure-Theoretic Definitions of Generalized Information
  Measures}
\label{Section:ME:MeasureTheoreticDefinitionsOfGeneralizedInformationMeasures}
        \noindent
%        In this section we extend the measure-theoretic definitions to
%        generalized information measures discussed in
%        Chapter~\ref{Chapter:KN}.
	We begin with a brief note on the notation and assumptions
        used. 
        We define all the information measures 
        on the measurable space $(X,\mathfrak{M})$, and default reference
        measure is $\mu$ unless otherwise stated. 
        To avoid clumsy formulations, we will not
        distinguish between functions differing on a $\mu$-null set
        only; nevertheless, we can work with equations between
        $\mathfrak{M}$-measurable functions on $X$ if they are
        stated as valid as being only $\mu$-almost everywhere ($\mu$-a.e or
        a.e).
        Further we assume that all the quantities of interest
        exist and assume, implicitly, the $\sigma$-finiteness of $\mu$ and
        $\mu$-continuity of probability measures whenever
        required. Since these assumptions repeatedly occur in various
        definitions and formulations, these will not be mentioned in
        the sequel.
        With these assumptions we do not distinguish between 
        an information measure of pdf $p$ and of corresponding probability
        measure $P$ -- hence we give definitions of
        information measures for pdfs, we use  corresponding
        definitions of probability measures as well, when ever it is
        convenient or required  --  with the understanding that $P(E) = \int_{E} p\,
        \ud \mu $, the converse being due to the Radon-Nikodym theorem, where $p =
        \frac{\ud P}{\ud \mu}$. In both the cases we have $P \ll \mu$.

        First we consider the R\'{e}nyi generalizations.
        Measure-theoretic definition of R\'{e}nyi entropy can be given
        as follows.
        %DEFINITION: Measure-theoretic definition of Renyi entropy
        \begin{definition}
        \label{Definition:ME:Measure-TheoreticRenyiEntropy}
        R\'{e}nyi entropy 
        of a pdf $p:X \rightarrow {\mathbb{R}}^{+}$ on a measure space
        $(X,\mathfrak{M},\mu)$ is defined as 
        \begin{equation}
        \label{Equation:ME:RenyiEntropyOf-pdf}  
        S_{\alpha}(p) = \frac{1}{1-\alpha} \ln 
        \int_{X}p(x)^{\alpha}\, \ud \mu(x) \enspace, 
        \end{equation}
        provided the integral on the right exists and $\alpha \in
        \mathbb{R}$, $\alpha > 0$.
        \end{definition}%EndDEFINITION: Measure-theoretic definition of RenyiEntropy
        The same can be defined for any $\mu$-continuous probability
        measure $P$ as
        \begin{equation}
        \label{Equation:ME:RenyiEntropyOf-PM}
          S_{\alpha}(P) = \frac{1}{1-\alpha} \ln  \int_{X}
          {\left( \frac{\ud P}{\ud \mu} \right)}^{\alpha -1} \, \ud P \enspace.
        \end{equation}  
        On the other hand, R\'{e}nyi relative-entropy can be defined as
        follows.
        %DEFINITION: Measure-theoretic definition of Tsallis relative entropy
        \begin{definition}
        Let $p,r:X \rightarrow
        {\mathbb{R}}^{+}$ be two pdfs on measure space $(X,\mathfrak{M},\mu)$. The
        R\'{e}nyi relative-entropy of $p$ relative to $r$ 
        is defined as
        \begin{equation}
        \label{Equation:ME:RenyiRelativeEntropyOf-pdf}              
        I_{\alpha}(p\|r) = \frac{1}{\alpha -1} \ln \int_{X}
        \frac{p(x)^{\alpha}}{r(x)^{\alpha -1}} \, \ud \mu(x) \enspace,
        \end{equation}
        provided the integral on the right exists and $\alpha \in
        \mathbb{R}$, $\alpha > 0$.
        \end{definition}%EndDEFINITION: Measure-theoretic definition of Tsallis
         %relative entropy
        The same can be written in terms of probability measures as,
        \begin{eqnarray}
	\label{Equation:ME:RenyiRelativeEntropyOf-PMs}
          I_{\alpha}(P\|R) &=& \frac{1}{\alpha -1} \ln   \int_{X}
          {\left( \frac{\ud P}{\ud R} \right)}^{\alpha -1} \, \ud P
          \nonumber \\
          &=& \frac{1}{\alpha -1} \ln   \int_{X}
          {\left( \frac{\ud P}{\ud R} \right)}^{\alpha} \, \ud R
          \enspace,
        \end{eqnarray}
	whenever $P \ll R$; $I_{\alpha}(P \|R) = + \infty$, otherwise.
	 Further if we assume $\mu$ in
        (\ref{Equation:ME:RenyiEntropyOf-PM}) is a probability measure
        then 
	\begin{equation}
	\label{Equation:ME:Renyi_EntropyandRelativeEntropy}
	S_{\alpha}(P) = I_{\alpha}(P\|\mu) \enspace.
	\end{equation}
        
        Tsallis entropy in the measure theoretic setting can be defined as
        follows.   
        %DEFINITION: Measure-theoretic definition of Tsallis entropy
        \begin{definition}
        \label{Definition:ME:Measure-TheoreticTsallisEntropy}
        Tsallis entropy of a pdf $p$ on $(X,\mathfrak{M},\mu)$ is
        defined as 
        \begin{equation}
        \label{Equation:ME:TsallisEntropyOf-pdf}  
        S_{q}(p) = \int_{X} p(x) \ln_{q} \frac{1}{p(x)}\, \ud \mu(x) =
        \frac{1 - \int_{X} p(x)^{q}\, \ud \mu(x) }{q-1}
        \enspace, 
        \end{equation}
        provided the integral on the right exists and $q \in
        \mathbb{R}$ and $q > 0$.
        \end{definition}%EndDEFINITION: Measure-theoretic definition
        %of TsallisEntropy

	$\ln_{q}$ in
            (\ref{Equation:ME:TsallisEntropyOf-pdf}) is referred to as
            $q$-logarithm and is defined as $\ln_{q} x = \frac{\displaystyle
            x^{1-q} -1}{\displaystyle 1-q} 
        \:\:\: (x >0, q \in {\mathbb{R}})$.
        The same can be defined for $\mu$-continuous probability
        measure $P$, and can be written as 
        \begin{equation}
	\label{Equation:ME:TsallisEntropyOf-PM}
           S_{q}(P) = \int_{X} \ln_{q}  {\left(\frac{\ud P}{\ud \mu}\right)}^{-1}
          \, \ud P \enspace.
        \end{equation}  
        
        The definition of Tsallis relative-entropy is given below. 
        %DEFINITION: Measure-theoretic definition of Tsallis relative entropy
        \begin{definition}
        Let $(X,\mathfrak{M},\mu)$ be a measure space. Let $p,r:X \rightarrow
        {\mathbb{R}}^{+}$ be two probability density functions. The
        Tsallis relative-entropy of $p$ relative to $r$ 
        is defined as
        \begin{equation}
        \label{Equation:ME:TsallisRelativeEntropyOf-pdf}            
        I_{q}(p\|r) = - \int_{X} p(x) \ln_{q} \frac{r(x)}{p(x)}\, \ud
        \mu(x)    = \frac{\int_{X} \frac{p(x)^{q}}{r(x)^{q-1}}\,
          \ud \mu(x) -1 }{q-1}
        \end{equation}
        provided the integral on right exists and $q \in
        \mathbb{R}$ and $q > 0$.
        \end{definition}%EndDEFINITION: Measure-theoretic definition of Tsallis
         %relative entropy
        The same can be written for two probability measures $P$ and
        $R$, as
        \begin{equation}
          I_{q}(P\|R)= - \int_{X} \ln_{q} {\left(\frac{\ud P}{\ud R}\right)}^{-1}\,
          \ud P \enspace,
        \end{equation}
	whenever $P \ll R$; $I_{q}(P \|R) = + \infty$, otherwise.
	If $\mu$ in
        (\ref{Equation:ME:TsallisEntropyOf-PM}) is a probability measure
        then 
	\begin{equation}
	\label{Equation:ME:Tsallis_EntropyandRelativeEntropy}
	S_{q}(P) = I_{q}(P\|\mu) \enspace.
	\end{equation}

%        We discuss the relations between generalized entropic
%        functionals in measure-theoretic case to discrete or continuous
%        case in
%        \S~\ref{Section:ME:MeasureTheoreticDefinitions_Revisited}. The
%        reason for this is the various relations discussed for
%        classical information measures cannot be extended to the
%        generalized case. As we are going to see contrary to the
%        classical case, where consistency of ME-prescriptions of measure-theoretic
%        definitions with discrete or continuous case can be argued
%        without invoking ME-prescriptions, consistent arguments for measure-theoretic
%        generalized entropy functionals involve explicitly
%        ME-prescriptions. Hence it is important for us to discuss the
%        ME-prescriptions in generalized case. First we briefly review
%        the ME-prescriptions in the classical case.

%=========================Section:=====================================
\section{Maximum Entropy and Canonical Distributions}
\label{Section:ME:MaximumEntropyAndCanonicalDistributions}
        \noindent
        For all the ME prescriptions of classical information measures
        we consider set of constrains of the form
        \begin{equation}
        \label{Equation:ME:ExpectationConstraints}
        \int_{X} u_{m} \, \ud P = \int_{X} u_{m}(x) p(x) \, \ud \mu(x) =
        \langle u_{m} \rangle \enspace, \:\:\:m = 
        1, \ldots , M \enspace,
        \end{equation}
        with respect to $\mathfrak{M}$-measurable functions $u_{m}: X
        \rightarrow \mathbb{R}, \:\: m = 1, \ldots M$, whose expectation
        values $\langle u_{m} \rangle, \, m=1,\ldots M$ are (assumed
        to be) {\it a priori} known, along with the normalizing
        constraint $\int_{X} \, \ud P =1$.
        (From now on we assume that any set of constraints on
        probability distributions implicitly includes this
        constraint, which will not be mentioned in the sequel.)

%-----Note on the notation for next chapter...	
%        A note on the notation: To avoid proliferation of symbols we
%        use the same notation for the minimum or maximum entropy
%        distributions and Lagrange multipliers in the various case;
%        the correspondence should be clear from the context. In the
%        maximum entropy case use $Z$ for the partition function and in
%        minimum entropy case we $\widehat{Z}$. 

        To maximize the
        entropy~(\ref{Equation:ME:ShannonEntropyOf-pdf})
        with respect
        to the constraints~(\ref{Equation:ME:ExpectationConstraints}), the
        solution is calculated via the Lagrangian:
        {\setlength\arraycolsep{0pt}
        \begin{eqnarray}
        \label{Equation:ME:LagranginForMaximumEntropy}
        \mathcal{L}(x, \lambda, \beta) = - \int_{X} \ln \frac{\ud
        P}{\ud \mu}(x)&& \, \ud P(x) - \lambda \left(\int_{X}\, \ud P(x) - 1
        \right) \nonumber \\
        && - \sum_{m=1}^{M} \beta_{m} \left(\int_{X} u_{m}(x)\, \ud P(x) -
        \langle u_{m} \rangle \right) \enspace, 
        \end{eqnarray}}
        where $\lambda$ and $\beta_{m}\, m=1,\ldots,M$ are Lagrange
        parameters (we use the notation $\beta = (\beta_{1}, \ldots, \beta_{M})$).
        \noindent
        The solution is given by
        \begin{displaymath}
        \ln \frac{\ud P}{\ud \mu}(x) + \lambda + \sum_{m=1}^{M}
        \beta_{m} u_{m}(x) = 0 \enspace.
        \end{displaymath}
        The solution can be calculated as  
        \begin{equation}
        \ud P(x, \beta) = \exp \left( -\ln Z(\beta) - \sum_{m=1}^{M}
        \beta_{m} u_{m}(x)\right) \ud \mu(x)
        \end{equation}
        or
        \begin{equation}
        p(x) = \frac{\ud P}{\ud \mu} (x) = \frac{e^{ -
            \sum_{m=1}^{M} \beta_{m} 
        u_{m}(x)}}{Z(\beta)}  \enspace,
        \end{equation}
        where the partition function $Z(\beta)$ is written as
        \begin{equation}
        \label{Equation:PartitionFunctionForMaximumEntropy}
        Z(\beta) = \int_{X} \exp \left( - \sum_{m=1}^{M} \beta_{m}
        u_{m}(x)\right) \ud \mu(x) \enspace.
        \end{equation}
        The Lagrange parameters $\beta_{m},\: m = 1, \ldots M$ are
        specified by the set of
        constraints (\ref{Equation:ME:ExpectationConstraints}).

        The maximum entropy, denoted by $S$, can be calculated as
        \begin{equation}
        \label{Equation:ME:MaximumEntropy}
        S = \ln Z + \sum_{m=1}^{M} \beta_{m} \langle u_{m} \rangle \enspace.
        \end{equation}

        The Lagrange parameters $\beta_{m},\: m = 1, \ldots M$, are
        calculated by searching the unique solution (if it exists) of the
        following system of nonlinear equations:
        \begin{equation}
        \label{Equation:ME:MaximumEntropy_ThermodynamicEquation_1}
          \frac{\partial}{\partial \beta_{m}} \ln Z(\beta) = - \langle
        u_{m} \rangle \enspace, \:\:\:m = 1, \ldots M \enspace. 
        \end{equation}
        We also have
        \begin{equation}
        \label{Equation:ME:MaximumEntropy_ThermodynamicEquation_2}      
        \frac{\partial S}{\partial \langle u_{m} \rangle} = -
        \beta_{m} \enspace, \:\:\: m = 1, \ldots M \enspace. 
        \end{equation}
        Equations
        (\ref{Equation:ME:MaximumEntropy_ThermodynamicEquation_1}) and 
        (\ref{Equation:ME:MaximumEntropy_ThermodynamicEquation_1}) are
        referred to as the thermodynamic equations.

%================================Section:===================================
\section{ME prescription for Tsallis Entropy}
\label{Section:ME:ME-prescriptionForTsallisEntropy}
        \noindent
         The great success of Tsallis entropy is
         attributed to the power-law distributions one can derive as
         maximum entropy distributions by maximizing Tsallis entropy
         with respect to the moment constraints. But there are
         subtilities  involved in the choice of constraints one would
         choose for ME prescriptions of these
         entropy functionals. These subtilities  are still part of the
         major discussion in the nonextensive formalism~\cite{FerriMartinezPlastino:2005:TheRoleOfConstraintsInTsallisNonextensiveTreatmentRevisited,AbeBagci:2005:NecessityOfqExpectation,WadaScarfone:2005:ConnectionsBetweenTsallisFormalismEtc}. 
         
        In the nonextensive formalism maximum entropy distributions
        are derived with respect to the constraints which are
        different from (\ref{Equation:ME:ExpectationConstraints}),
        which are used for classical information measures. The
        constraints of the
        form~(\ref{Equation:ME:ExpectationConstraints}) are
        inadequate for handling the serious mathematical difficulties
        (see~\cite{TsallisMendesPlastino:1998:TheRoleOfConstraints}). To
        handle these difficulties constraints of the form
        \begin{equation}
        \label{Equation:ME:Normalized-q-ExpectationConstraints}  
        \frac{\int_{X} u_{m}(x) p(x)^{q} \, \ud \mu(x)}{\int_{X}
          p(x)^{q}\, \ud \mu(x)} = {\langle\langle u_{m} \rangle\rangle}_{q} \enspace, m = 
        1, \ldots , M
        \end{equation}
	are proposed.
        (\ref{Equation:ME:Normalized-q-ExpectationConstraints}) can
          be considered as the expectation with respect to the
          modified probability measure $P_{(q)}$ (it is indeed a
          probability measure) defined as
          \begin{equation}
            P_{(q)}(E) = {\left( \int_{X} p(x)^{q} \, \ud \mu
              \right)}^{-1} \int_{E} p(x)^{q} \, \ud \mu \enspace.
          \end{equation}
          The measure $ P_{(q)}$ is known as escort probability
          measure. 

          The variational principle for Tsallis entropy maximization
          with respect to
          constraints~(\ref{Equation:ME:Normalized-q-ExpectationConstraints})
          can be written as
          \begin{eqnarray}
          \label{Equation:ME:Lagrangin_TsallisMaximumEntropy_wrt_Norm-q-Expt}
          \mathcal{L}(x, \lambda, \beta) =  &&\int_{X} \ln_{q}
          \frac{1}{p(x)} \, \ud P(x) - \lambda \left(\int_{X}\, \ud P(x) - 1
          \right) \nonumber \\
          && - \sum_{m=1}^{M} \beta^{(q)}_{m} \left(\int_{X} {p(x)}^{q-1}
          \left(u_{m}(x) - {\langle\langle u_{m}  \rangle\rangle}_{q}
          \right) \, \ud P(x) \right) \enspace,
          \end{eqnarray}
          where the parameters $\beta_{m}^{(q)}$ can be defined in
          terms of true Lagrange parameters $\beta_{m}$ as
         \begin{equation}
           \beta_{m}^{(q)} = {\left(\int_{X} p(x)^{q}\, \ud \mu
             \right)}^{-1} \beta_{m}\enspace, \, m = 1, \ldots, M.
          \end{equation}
          The maximum entropy distribution in this case can be written
          as
          \begin{equation}
          \label{Equation:ME:TsallisMaximumEntropyDistribution_wrt_q-Expt}   
          p(x) = \frac{\displaystyle {\left[ 1 - (1-q)  {\left( \int
            dx\,{p(x)}^{q} \right)}^{-1}  \sum_{m=1}^{M} \beta_{m} \left( u_{m}(x) - 
          {\langle\langle {u}_{m} \rangle\rangle}_{q} \right) \right]}^{\frac{1}{1-q}}}
          {\displaystyle {\overline{Z_{q}}}  }
          \end{equation}

         \begin{equation}
         \label{Equation:ME:TsallisMaximumEntropyDistribution_wrt_q-Expt_q-exponentialForm} 
         p(x) = \frac{\displaystyle e_{q}^{-   {\left(\int_{X} p(x)^{q}\, \ud \mu
             \right)}^{-1}   \sum_{m=1}^{M} \beta_{m} (u_{m}(x) -
             {\langle\langle u_{m}\rangle\rangle}_{q}  )
         }}{\displaystyle \overline{Z_{q}}} \enspace,         
         \end{equation}
         where
         \begin{equation}
           \overline{Z_{q}} = \int_{X} {e_{q}^{- {\left(\int_{X} p(x)^{q}\, \ud \mu
             \right)}^{-1}   \sum_{m=1}^{M} \beta_{m} (u_{m}(x) -
             {\langle\langle u_{m}\rangle\rangle}_{q}  ) }} \, \ud \mu(x) \enspace.
         \end{equation}

        Maximum Tsallis entropy in this case satisfies
        \begin{equation}
        S_{q} = \ln_{q}\overline{{Z}_{q}} \enspace,
        \end{equation}
        while corresponding thermodynamic equations can be written
        as 
        \begin{equation}
        \frac{\partial}{\partial \beta_{m}} \ln_{q} Z_{q}  =  -
        {\langle\langle{{u}_{m}}\rangle\rangle}_{q} \enspace, \:\:\: m = 1, \ldots M
        \enspace,
        \end{equation}
        \begin{equation}
        \frac{\partial S_{q}}{\partial
        {\langle\langle{{u}_{m}}\rangle\rangle}_{q}  }  =  -
        \beta_{m} \enspace, \:\:\: m =1, \ldots M \enspace,
        \end{equation}
        where
        \begin{equation}
        \ln_{q} Z_{q} = \ln_{q} \overline{{Z}_{q}}
        - \sum_{m=1}^{M} \beta_{m}
        {\langle\langle{{u}_{m}}\rangle\rangle}_{q} \enspace.
        \end{equation}

%=============================================================================
\section{Measure-Theoretic Definitions: Revisited}
\label{Section:ME:MeasureTheoreticDefinitions_Revisited}
       \noindent
	It is well known that unlike Shannon entropy, Kullback-Leibler
       relative-entropy in the discrete 
       case can be extended naturally to the measure-theoretic
       case. 
       In this section, we show
       that this fact is true for generalized relative-entropies
       too. R\'{e}nyi relative-entropy on continuous valued space
       $\mathbb{R}$ and its  
       equivalence with the discrete case is studied
       by R\'{e}nyi~\cite{Renyi:1960:SomeFundamentalQuestionsOfInformationTheory}. Here,
       we present the result in the measure-theoretic case and
       conclude that both measure-theoretic definitions of Tsallis and
       R\'{e}nyi relative-entropies are equivalent to its discrete
       case. 

       We also present a result pertaining to ME of
       measure-theoretic Tsallis entropy. We show that ME of Tsallis
       entropy in the measure-theoretic case is consistent with the
       discrete case.

   %-----------------------Sub Section------------------     
  \subsection{On Measure-Theoretic Definitions of Generalized Relative-Entropies}
       \noindent
        Here we show that generalized relative-entropies in the
        discrete case can be naturally extended to measure-theoretic
        case, in the  sense that measure-theoretic definitions can
        be defined as a limit of a sequence of finite discrete
        entropies of pmfs which approximate the pdfs involved.
        We call this
        sequence of pmfs as ``approximating sequence of pmfs of a
        pdf''. To formalize these aspects we need the following 
        lemma. 
        %--------------Lemmma-------------
        \begin{lemma}
        \label{Lemma:ME:ExistenceOfApproximatingSequenceOfSimpleFunctionsForPdf}  
        Let $p$ be a pdf defined on measure space
        $(X,\mathfrak{M},\mu)$. Then there exists a sequence of simple
        functions $\{f_{n}\}$ (we refer to them as approximating sequence of
        simple functions of $p$) such that $\lim_{n \to \infty} f_{n} = p$
        and each $f_{n}$ can be written as
        \begin{equation}
        \label{Equation:ME:ActualDefinitionOfSeqenceOfSimpleFunctions} 
         f_{n}(x) = \frac{1}{\mu(E_{n,k})} \int_{E_{n,k}} p \, \ud 
        \mu \enspace, \:\:\:\:\:\:\: \forall x \in E_{n,k},
	 \:\:\: k = 1, \ldots m(n) \enspace,
        \end{equation}  
        where $(E_{n,1}, \ldots, E_{n,m(n)})$ is the measurable 
        partition corresponding to $f_{n}$ (the notation $m(n)$
        indicates that $m$ varies with $n$).  Further each $f_{n}$
  	satisfies 
        \begin{equation}
         \int_{X} f_{n} \, \ud \mu = 1 \enspace.
        \end{equation}  
        \end{lemma}
        %Proof----
        \proof
%	\footnote{$ \cup_{k=1}^{m(n)} E_{n,k} = X$ and $E_{n,i}
%        \cap E_{n,j} = \emptyset$, $\forall i \neq j$} 
         Define a sequence of simple functions $\{f_{n}\}$ as
        \begin{equation}
         f_{n}(x) = \left\{ \begin{array}{ll}
          \frac{1}{ \mu p^{-1} \left(
            \left[ \frac{k}{2^{n}}, \frac{k+1}{2^{n}} \right) \right)}
            \displaystyle \int_{  p^{-1} \left(
            \left[ \frac{k}{2^{n}}, \frac{k+1}{2^{n}} \right) \right)
            } p \, \ud \mu \enspace,& \: \:
         \:\:\textrm{if}\:\:  \frac{k}{2^{n}} \leq p(x) <
         \frac{k+1}{2^{n}} , \\
         & \:\:\:k = 0, 1, \ldots n 2^{n}-1
         \\ \\
         \frac{1}{ \mu p^{-1} \left(
            \left[ n, \infty \right) \right)}
            \displaystyle \int_{  p^{-1} \left(
            \left[ n , \infty \right) \right)
            } p \, \ud \mu \enspace,& \: \:
         \:\:\textrm{if}\:\: n \leq p(x),
           \end{array} \right.
         \end{equation}
         Each $f_{n}$ is indeed a simple function and can be written as
         \begin{equation}
          f_{n} = \sum_{k=0}^{n2^{n}-1} \left( \frac{1}{\mu E_{n,k}} 
          \int_{E_{n,k}} p\, \ud \mu \right) \chi_{E_{n,k}} + \left( \frac{1}{\mu
            F_{n}} \int_{F_{n}} p \, \ud \mu \right) \chi_{F_{n}} \enspace, 
         \end{equation}
         where $E_{n,k} =
         p^{-1}\left(\left[\frac{k}{2^{n}},\frac{k+1}{2^{n}}\right)
          \right)$, $k= 0, \ldots, n2^{n}-1$ and $F_{n} = p^{-1} \left(
            \left[ n, \infty \right) \right)$. 
         Since $\int_{E} p \, \ud \mu < \infty$ for any $E \in
         \mathfrak{M}$, we have $\int_{E_{n,k}} p\, \ud \mu = 0$
         whenever $\mu E_{n,k} =0$, for $k = 0, \ldots n2^{n} -1$. Similarly 
         $\int_{F_{n}} p\, \ud \mu = 0$ whenever $\mu F_{n} =0$.
         Now we show that $\lim_{n \to \infty} f_{n} = p$, point-wise.
         
         First assume that $p(x) < \infty$. Then $\exists \: n \in
         {\mathbb{Z}}^{+} \ni p(x) \leq n$. Also $\exists \, k \in
         {\mathbb{Z}}^{+} $, $0 \leq k 
         \leq n2^{n}-1
         \ni \frac{k}{2^{n}} \leq p(x) < 
         \frac{k+1}{2^{n}}$ and $\frac{k}{2^{n}} \leq f_{n}(x) <
         \frac{k+1}{2^{n}}$. This implies $0 \leq |p - f_{n} | <
         \frac{1}{2^{n}}$ as required.

         If $p(x) = \infty$, for some $x \in X$, then $x \in F_{n}$ for
         all $n$, and therefore $f_{n}(x) \geq n$ for all $n$; hence
         $\lim_{n \to \infty} f_{n}(x) = \infty = p(x) $.

         Finally we have
         \begin{eqnarray}
           \int_{X} f_{n} \, \ud \mu &=& \sum_{k=1}^{n(m)} \left[
           \frac{1}{\mu(E_{n,k})} \int_{E_{n,k}} p \,\ud \mu \right]
            \mu(E_{n,k}) \nonumber \\
            &=& \sum_{k=1}^{n(m)} \int_{E_{n,k}} p \,\ud \mu \nonumber \\
            &=& \int_{X} p \, \ud \mu =1 \nonumber
         \end{eqnarray}  
         \endproof
         %-------------End: lemmma-----------------
         The above construction of a sequence of simple functions which
         approximate a measurable function is similar to the
         approximation theorem~\cite[pp.6, Theorem
           1.8(b)]{Kantorovitz:2003:IntroductionToModernAnalysis} in
         the theory of integration. But, approximation in
         Lemma~\ref{Lemma:ME:ExistenceOfApproximatingSequenceOfSimpleFunctionsForPdf}
         can be seen as a mean-value approximation where as in the later
         case it is the lower approximation. Further, unlike in the case
         of lower approximation, the sequence of simple functions 
         which approximate $p$ in
         Lemma~\ref{Lemma:ME:ExistenceOfApproximatingSequenceOfSimpleFunctionsForPdf}
         are neither monotone nor satisfy $f_{n} \leq p$.
             
        Now one can define a sequence of pmfs $\{\tilde{p}_{n}\}$ corresponding
        to the sequence 
        of simple functions constructed in
        Lemma~\ref{Lemma:ME:ExistenceOfApproximatingSequenceOfSimpleFunctionsForPdf},
        denoted by $\tilde{p}_{n} = (\tilde{p}_{n,1}, \ldots,\tilde{p}_{n,m(n)})$, as 
        \begin{equation}
        \label{Equation:ME:ActualDefinitionOfSeqenceOfPmfs} 
         \tilde{p}_{n,k} = \mu(E_{n,k})f_{n}\chi_{E_{n,k}} = \int_{E_{n,k}} p \, \ud 
         \mu \enspace, k = 1, \ldots m(n),
        \end{equation}
        for any $n$.
        We have
        \begin{equation}
         \sum_{k=1}^{m(n)} \tilde{p}_{n,k} = \sum_{k=1}^{m(n)} \int_{E_{n,k}} p
         \, \ud \mu
         = \int_{X} p \, \ud \mu =1 \enspace,
        \end{equation}
        and hence $\tilde{p}_{n}$ is indeed a pmf. 
        We call $\{\tilde{p}_{n}\}$ as the approximating sequence of pmfs of pdf
        $p$.
       
%        We say an measure-theoretic definition of an information
%        measure $\overline{S}$ is exact if
%        \begin{equation}
%         \lim_{n \to \infty} \overline{S}(P_{n}) = \overline{S}(p) \enspace.
%        \end{equation}

          Now we present our main theorem, where we assume that $p$ and
	  $r$ are bounded. The
          assumption of boundedness of $p$ and $r$ simplifies the
	  proof. However, the result can be
          extended to an unbounded
          case. See~\cite{Renyi:1959:OnTheDimensionAndEntropyOfProbabilityDistributions}
          analysis of Shannon entropy and relative entropy on $\mathbb{R}$.
         %THEOREM:Measure-theoretic definition of generalized relative entropies.
         \begin{theorem}
         \label{Theorem:ME:MeasureTheoreticDefinitionsOfGeneralizedRelative-Entropies}
            Let $p$ and $r$ be pdf, which are bounded, defined on a
            measure space $(X,\mathfrak{M}, \mu)$. Let $\tilde{p}_{n}$
            and $\tilde{r}_{n}$ be the approximating sequence of pmfs of $p$ and $r$
            respectively. Let $I_{\alpha}$ denotes the R\'{e}nyi relative-entropy as
            in~(\ref{Equation:ME:RenyiRelativeEntropyOf-pdf}) and
	  $I_{q}$ denote the Tsallis 
            relative-entropy as
            in~(\ref{Equation:ME:TsallisRelativeEntropyOf-pdf}) 
            then
            \begin{equation}
            \label{Equation:ME:InRenyisTheoremStatement_2}              
            \lim_{n \to \infty} I_{\alpha}(\tilde{p}_{n} \| \tilde{r}_{n}) = I_{\alpha}(p\|r)
            \end{equation}
	     and
            \begin{equation}
            \label{Equation:ME:InRenyisTheoremStatement_1}  
            \lim_{n \to \infty} I_{q}(\tilde{p}_{n} \| \tilde{r}_{n}) = I_{q}(p\|r)
            \end{equation}
         \end{theorem}  
         \proof
         It is enough to prove the result for either Tsallis or
         R\'{e}nyi since each are monotone and continuous functions of
         each other. Hence we write down the proof for the case of R\'{e}nyi
         and we use the entropic index $\alpha$ in the proof.

         Corresponding to pdf $p$, let $\{f_{n}\}$ be the approximating 
         sequence of simple functions such that $\lim_{n \to \infty}
         f_{n} = p$ as in
         Lemma~\ref{Lemma:ME:ExistenceOfApproximatingSequenceOfSimpleFunctionsForPdf}.
         Let $\{g_{n}$ be the approximating sequence of simple
         functions for $r$ such that $\lim_{n \to \infty} g_{n} = r$.  
         Corresponding
         to simple functions $f_{n}$ and $g_{n}$ there exists a common
         measurable partition\footnote{Let $\varphi$ and $\phi$ are two
         simple functions defined on $(X,\mathfrak{M})$. Let $\{E_{1},
         \ldots E_{n}\}$ and $\{F_{1},\ldots, F_{m}\}$ be the measurable
         partitions corresponding to $\varphi$ and $\phi$
         respectively. Then partition defined as $\{E_{i} \cap E_{j} |
         i = 1, \ldots n,\:\: j =1, \ldots m\}$ is a common measurable
         partition for both $\varphi$ and $\phi$.}
         $\{ E_{n,1}, \ldots E_{n,m(n)}\}$ such
         that $f_{n}$ and $g_{n}$ can be written as
         \begin{equation}
         \label{Equation:ME:InRenyisTheorem_1_a}  
           f_{n}(x) = \sum_{k=1}^{m(n)} (a_{n,k})
           \chi_{E_{n,k}}(x) \enspace, \:\:\: a_{n,k} \in
               {\mathbb{R}}^{+}, \, \forall k = 1, \ldots m(n) \enspace,
         \end{equation}
         \begin{equation}
         \label{Equation:ME:InRenyisTheorem_1_b}             
           g_{n}(x) = \sum_{k=1}^{m(n)} (b_{n,k})
           \chi_{E_{n,k}}(x) \enspace, \:\:\: b_{n,k} \in
               {\mathbb{R}}^{+}, \, \forall k = 1, \ldots m(n) \enspace,
         \end{equation}
         where  $\chi_{E_{n,k}}$ is the characteristic function of
         $E_{n,k}$, for $k=1,\ldots m(n)$. By
         (\ref{Equation:ME:InRenyisTheorem_1_a}) and
         (\ref{Equation:ME:InRenyisTheorem_1_b}) the approximating
         sequences of pmfs $\{\tilde{p}_{n} = (\tilde{p}_{n,1},
         \ldots, \tilde{p}_{n,m(n)})\}$  
          and $\{\tilde{r}_{n} = (\tilde{r}_{n,1}, \ldots,
         \tilde{r}_{n,m(n)})\}$ can be written as
%	corresponding 
%          to pdfs $p$ and $r$ respectively can be written as $\tilde{p}_{n,k}
%          =  (a_{n,k}) \mu(E_{n,k}),\, k = 1, \ldots , m(n) $ and
%          $ \tilde{r}_{n,k}= (b_{n,k}) \mu(E_{n,k}), \, k = 1, \ldots ,
%          m(n)$
          (see (\ref{Equation:ME:ActualDefinitionOfSeqenceOfPmfs}))
         \begin{equation}
	 \label{Equation:ME:InRenyisTheorem_2_a}
           \tilde{p}_{n,k} =  a_{n,k} \mu(E_{n,k})\:\:\: k = 1, \ldots , m(n) \enspace,
         \end{equation}
         \begin{equation}
	 \label{Equation:ME:InRenyisTheorem_2_b}
           \tilde{r}_{n,k} =  b_{n,k} \mu(E_{n,k})\:\:\: k = 1, \ldots , m(n) \enspace.
         \end{equation}
      	 Now R\'{e}nyi
         relative entropy for $\tilde{p}_{n}$ and 
         $\tilde{r}_{n}$ can be written as
         \begin{equation}
         \label{Equation:ME:InRenyisTheorem_2}  
           S_{\alpha}(\tilde{p}_{n} \| \tilde{r}_{n}) =
         \frac{1}{\alpha-1} \ln \sum_{k=1}^{m(n)} 
           \frac{a_{n,k}^{\alpha}}{b_{n,k}^{\alpha -1}}
           \mu(E_{n,k}) \enspace.
         \end{equation}

         To prove $\lim_{n \rightarrow \infty} S_{\alpha}(\tilde{p}_{n} \|
         \tilde{r}_{n}) = S_{\alpha}(p \| r) $ it is enough to prove that
         \begin{equation}
         \label{Equation:ME:InRenyisTheorem_2}             
           \lim_{n \rightarrow \infty} \frac{1}{\alpha-1} \ln
           \int_{X} \frac{ {f_{n}(x)}^{\alpha} }{
             {g_{n}(x)}^{\alpha-1}} \, \ud \mu(x)
            =   \frac{1}{\alpha-1} \ln
           \int_{X} \frac{ {p(x)}^{\alpha} }{
             {r(x)}^{\alpha-1}} \, \ud \mu(x) \enspace,
          \end{equation}  
           since we have\footnote{ Since simple functions
           ${\left(f_{n}\right)}^{\alpha}$ and ${\left(g_{n}\right)}^{\alpha-1}$ can be
           written as
           \begin{displaymath}
             {\left(f_{n}\right)}^{\alpha}(x) = \sum_{k=1}^{m(n)} 
             \left( a_{n,k}^{\alpha} \right) \chi_{E_{n,k}}(x)
            \enspace, \:\:\:\:\:\mbox{and}
           \end{displaymath}
           \begin{displaymath}
             {\left(g_{n}\right)}^{\alpha-1}(x) = \sum_{k=1}^{m(n)} 
             \left( b_{n,k}^{\alpha-1} \right) \chi_{E_{n,k}}(x) \enspace.
           \end{displaymath}
           Further,
           \begin{displaymath}
           \frac{f_{n}^{\alpha}}{g_{n}^{\alpha-1}}(x)
            =    \sum_{k=1}^{m(n)} \left( \frac{ 
             a_{n,k}^{\alpha} }{b_{n,k}^{\alpha-1}} \right)   \chi_{E_{n,k}}(x) \enspace.
           \end{displaymath}
         }%Endfootnote 
         \begin{equation}
         \label{Equation:ME:InRenyisTheorem_3}              
           \int_{X} \frac{{f_{n}(x)}^{\alpha}}{{g_{n}(x)}^{\alpha -1} } \,
           \ud \mu(x) =
           \sum_{k=1}^{m(n)}
           \frac{a_{n,k}^{\alpha}}{b_{n,k}^{\alpha-1}} \mu(E_{n,k}) \enspace.  
         \end{equation}  
          Further it is enough  to prove that 
         \begin{equation}
         \label{Equation:ME:InRenyisTheorem_3}                         
           \lim_{n \rightarrow \infty} 
           \int_{X}  {h_{n}(x)}^{\alpha} g_{n}(x)   \, \ud \mu(x)
            =   
           \int_{X} \frac{{p(x)}^{\alpha} }{
             {r(x)}^{\alpha-1}} \, \ud \mu(x) \enspace,
          \end{equation}
         where $h_{n}$ is defined as $h_{n}(x) =
         \frac{f_{n}(x)}{g_{n}(x)} $.\\ 
        %Case 1---------
        \noindent
        {\em \underline {Case 1: $0 < \alpha < 1$}}
        
        In this case
        the {\em Lebesgue dominated convergence
          theorem}~\cite[pp.26]{Rudin:1966:RealAndComplexAnalysis}
        gives that,
         \begin{equation}
         \label{Equation:ME:InRenyisTheorem_4}                                    
           \lim_{n \to \infty} \int_{X}
           \frac{f_{n}^{\alpha}}{g_{n}^{\alpha -1}} \, \ud \mu =
           \int_{X} \frac{p^{\alpha}}{r^{\alpha -1}} \, \ud \mu \enspace.
         \end{equation}
         and hence (\ref{Equation:ME:InRenyisTheoremStatement_1})

         %Case 2-----------
         \noindent
         {\em \underline {Case 2: $\alpha  > 1$}}

         We have $h_{n}^{\alpha} f_{n}
         \rightarrow \frac{f(x)^{\alpha}}{g(x)^{\alpha-1}}$ {\em
           a.e}. By {\em Fatou's
           Lemma}~\cite[pp.23]{Rudin:1966:RealAndComplexAnalysis} we
         obtain that, 
         \begin{equation}
         \label{Equation:ME:InRenyisTheorem_LimInfInequality}  
           \lim_{n \to \infty} \inf \int_{X}
           h_{n}(x)^{\alpha} g_{n}(x) \, \ud \mu(x) \geq
           \int_{X} \frac{{p(x)}^{\alpha} }{
             {r(x)}^{\alpha-1}} \, \ud \mu(x) \enspace.
         \end{equation}
         From the construction of $f_{n}$ and $g_{n}$
         (Lemma~\ref{Lemma:ME:ExistenceOfApproximatingSequenceOfSimpleFunctionsForPdf}) 
         we have
         \begin{equation}
         \label{Equation:ME:InRenyisTheorem_5}                                               
         h_{n}(x) f_{n}(x) = \frac{1}{\mu(E_{n,i})} \int_{E_{n,i}} 
         \frac{p(x)}{r(x)} r(x) \, \ud \mu \enspace, \:\:\: \forall x
         \in E_{n,i} \enspace.
         \end{equation}
         By Jensen's inequality we get
         \begin{equation}
         \label{Equation:ME:InRenyisTheorem_6}
         h_{n}(x)^{\alpha} f_{n}(x) \leq \frac{1}{\mu(E_{n,i})}
           \int_{E_{n,i}}  \frac{p(x)^{\alpha}}{r(x)^{\alpha-1}}   \,
           \ud \mu \enspace, \:\:\: \forall x \in E_{n,i} \enspace.
         \end{equation}  
         By (\ref{Equation:ME:InRenyisTheorem_1_a}) and
         (\ref{Equation:ME:InRenyisTheorem_1_b}) we can write
         (\ref{Equation:ME:InRenyisTheorem_6}) as
         \begin{equation}
         \label{Equation:ME:InRenyisTheorem_7}           
           \frac{a_{n,i}^{\alpha}}{b_{n,i}^{\alpha-1}}  \mu(E_{n,i})   \leq
           \int_{E_{n,i}}   \frac{p(x)^{\alpha}}{r(x)^{\alpha-1}}
           \, \ud \mu \enspace, \:\:\: \forall i = 1, \ldots m(n) \enspace.
         \end{equation}  
         By taking summations both sides of
         (\ref{Equation:ME:InRenyisTheorem_7}) we get 
         \begin{equation}
         \label{Equation:ME:InRenyisTheorem_8}                      
          \sum_{i=1}^{m(n)}  \frac{a_{n,i}^{\alpha}}{b_{n,i}^{\alpha-1}}  \mu(E_{n,i})   \leq
          \sum_{i=1}^{m(n)} \int_{E_{n,i}}
          \frac{p(x)^{\alpha}}{r(x)^{\alpha-1}}  \, \ud \mu \enspace,
          \:\:\: \forall i = 1, \ldots m(n) \enspace.
         \end{equation}
         The above equation (\ref{Equation:ME:InRenyisTheorem_8}) nothing but
         \begin{displaymath}
          \int_{X} h_{n}^{\alpha}(x) f_{n}(x) \, \mu(x)   \leq
          \int_{X}  \frac{p(x)^{\alpha}}{r(x)^{\alpha-1}}
            \, \ud \mu \enspace, \:\:\: \forall n \enspace,
         \end{displaymath}
         and hence
         \begin{displaymath}
         \sup_{i > n } \int_{X} h_{i}^{\alpha}(x) f_{i}(x) \, \mu(x)
         \leq \int_{X}  \frac{p(x)^{\alpha}}{r(x)^{\alpha-1}} 
           \, \ud \mu \enspace, \:\:\: \forall n \enspace.
         \end{displaymath}
         Finally we have
         \begin{equation}
         \label{Equation:ME:InRenyisTheorem_LimSupInequality}  
           \lim_{n \to \infty} \sup \int_{X}
           h_{n}^{\alpha}(x) f_{n}(x) \, \mu(x)   \leq \int_{X}
            \frac{p(x)^{\alpha}}{r(x)^{\alpha-1}} \, \ud \mu \enspace.
         \end{equation}
         From (\ref{Equation:ME:InRenyisTheorem_LimInfInequality}) and
         (\ref{Equation:ME:InRenyisTheorem_LimSupInequality}) we have
         \begin{equation}
         \label{Equation:ME:InRenyisTheorem_LimEquality}  
           \lim_{n \to \infty} \int_{X}
             \frac{f_{n}(x)^{\alpha}}{g_{n}(x)^{\alpha-1}}  \, \mu(x) = \int_{X}
            \frac{p(x)^{\alpha}}{r(x)^{\alpha-1}} \, \ud \mu \enspace,
         \end{equation}
         and hence (\ref{Equation:ME:InRenyisTheoremStatement_1}).
         \endproof 
          %EndProof--------------
         
  %--------------------Sub Section-----------------------------
  \subsection{On ME of Measure-Theoretic  definition of Tsallis entropy}
         \noindent
         With the shortcomings of Shannon entropy that it cannot be
         naturally extended to the non-discrete case, we have observed
         that Shannon entropy in its general case on measure space can
         be used consistently for the ME-prescriptions. One can easily
         see that generalized information measures of R\'{e}nyi and Tsallis
         too cannot be extended naturally to measure-theoretic case,
         i.e., measure-theoretic definitions are not equivalent to the 
         discrete case in the sense that they can not be defined as a
         limit of sequence of finite discrete entropies corresponding to
         pmfs defined on measurable partitions which approximates the
         pdf. One can use the same counter example we discussed in
         \S~\ref{SubSection:ME:DiscreteToContinuous}. We have already
         given the ME-prescriptions of Tsallis entropy in the
         measure-theoretic case. In this section, we show that the
         ME-prescriptions in the measure-theoretic case are consistent 
         with the discrete case.

 	 Proceeding as in the case of measure-theoretic entropy in
         \S~\ref{SubSection:ME:MeasureTheoreticCasesinDiscrete},
         measure-theoretic Tsallis
         entropy $S_{q}(P)$~(\ref{Equation:ME:TsallisEntropyOf-PM}) in
         the discrete case can be written as
         \begin{equation}
	 \label{Equation:ME:MeasureTheoreticTsallisEntropyInDiscreteForm}
         S_{q}(P) = \sum_{k=1}^{n} P_{k} \ln_{q} \frac{\mu_{k}}{P_{k}} \enspace.
         \end{equation}
         By (\ref{Equation:KN:PropertyOflnq(x/y)}) we get
         \begin{equation}
	 \label{Equation:ME:MeasureTheoreticTsallisEntropyInDiscreteForm_1}
         S_{q}(P) = \sum_{k=1}^{n} P_{k}^{q} \left[ \ln_{q} \mu_{k} -
         \ln_{q} P_{k} \right] = S_{q}^{n}(P) + \sum_{k=1}^{n} P_{k}^{q}
         \ln_{q} \mu_{k} \enspace,
         \end{equation}
         where $S_{q}^{n}(P)$ is the Tsallis entropy in discrete case.
         When $\mu$ is a uniform distribution i.e., $\mu_{k} =
         \frac{1}{n}\:\: \forall n = 1, \ldots n$ we get
         \begin{equation}
	 \label{Equation:ME:MeasureTheoreticTsallisEntropyInDiscreteForm_1}
         S_{q}(P) = S_{q}^{n}(P) - n^{q-1} \ln_{q} n \sum_{k=1}^{n}
         P_{k}^{q} \enspace.
         \end{equation}
         Now we show that the quantity $\sum_{k=1}^{n} P_{k}^{q}$ is
         constant in maximization of $S_{q}(P)$ with respect to the
         set of constraints
         (\ref{Equation:ME:Normalized-q-ExpectationConstraints}).

         The claim is that
         \begin{equation}
         \label{Equation:ME:SumOfpPowerqs_ForNormalizedExpectation}
         \int p(x)^{q}\, \ud \mu(x) = {(\overline{Z_{q}})}^{1-q} \enspace,
         \end{equation}
	 which holds for Tsallis maximum entropy distribution
         (\ref{Equation:ME:TsallisMaximumEntropyDistribution_wrt_q-Expt})
         in general. This can be shown as follows. From
         the maximum entropy 
         distribution~(\ref{Equation:ME:TsallisMaximumEntropyDistribution_wrt_q-Expt}),
         we have 
         \begin{displaymath}
         p(x)^{1-q} = \frac{\displaystyle 1 - (1-q)  {\left( \int_{X}
            {p(x)}^{q}\, \ud \mu(x) \right)}^{-1}  \sum_{m=1}^{M}
         \beta_{m} \left( u_{m}(x) -  
         {\langle\langle {u}_{m} \rangle\rangle}_{q} \right)}
         {\displaystyle ({\overline{Z_{q}}})^{1-q}  } \enspace,
         \end{displaymath}
         which can be rearranged as
         \begin{displaymath}
         ({\overline{Z_{q}}})^{1-q} p(x) = \left[ 1 - (1-q)
         \frac{\sum_{m=1}^{M} \beta_{m} \left( u_{m}(x) -  
         {\langle\langle {u}_{m} \rangle\rangle}_{q} \right)}{\int
         {p(x)}^{q}} \, \ud \mu(x) \right] p(x)^{q} \enspace.
         \end{displaymath}
         By integrating both sides in the above equation, and by
         using~(\ref{Equation:ME:Normalized-q-ExpectationConstraints})
         we get (\ref{Equation:ME:SumOfpPowerqs_ForNormalizedExpectation}).

         Now, (\ref{Equation:ME:SumOfpPowerqs_ForNormalizedExpectation}) can
         be written in its discrete form as
         \begin{equation}
         \label{Equation:ME:SumOfpPowerqs_ForNormalizedExpectation_Discrete_1}
          \sum_{k=1}^{n} \frac{P_{k}^{q}}{\mu_{k}^{q-1}} =
         {(\overline{Z_{q}})}^{1-q} \enspace.
         \end{equation}
         When $\mu$ is uniform distribution we get
         \begin{equation}
         \label{Equation:ME:SumOfpPowerqs_ForNormalizedExpectation_Discrete_2}
          \sum_{k=1}^{n} P_{k}^{q} = n^{1-q}  {(\overline{Z_{q}})}^{1-q}
         \end{equation}
         which is a constant.

         Hence by
         (\ref{Equation:ME:MeasureTheoreticTsallisEntropyInDiscreteForm_1})
         and
         (\ref{Equation:ME:SumOfpPowerqs_ForNormalizedExpectation_Discrete_2}),
         on can conclude that with respect to a particular instance of
         ME, measure-theoretic Tsallis entropy $S(P)$ defined for a
         probability measure $P$ on 
        the measure space $(X,\mathfrak{M},\mu)$, is equal to 
        discrete Tsallis entropy up to an 
        additive constant, when the reference measure $\mu$ is chosen as a uniform
        probability distribution. There by, one can further conclude
         that with respect to a particular instance of ME of
         measure-theoretic Tsallis entropy is consistent with its 
         discrete definition. 

%=======================Section: Conclusition===================
\section{Conclusions}
\label{Section:Conclusions}
	\noindent
	In this paper we presented measure-theoretic definitions of
	generalized information measures. We proved that the measure-theoretic
        definitions of generalized relative-entropies, R\'{e}nyi and
        Tsallis, are natural extensions of their respective discrete
        cases. We also showed that, ME prescriptions of
        measure-theoretic Tsallis entropy are consistent with the
        discrete case.

%========================Bibliography===================================
\section*{References}

\bibliographystyle{unsrt}
\bibliography{papi}

\begin{thebibliography}{10}

\bibitem{ShannonWeawer:1949:TheMathematicalTheoryOfCommunication}
C.~E. Shannon and W.~Weawer.
\newblock {\em The Mathematical Theory of Communication}.
\newblock University of Illinois Press, Urbana, Illinois, 1949.

\bibitem{Ash:1965:InformationTheory}
R.~B. Ash.
\newblock {\em Information Theory}.
\newblock Interscience, New York, 1965.

\bibitem{KapurKesavan:1997:EntropyOptimizationPrinciples}
J.~N. Kapur and H.~K. Kesavan.
\newblock {\em Entropy Optimization Principles with Applications}.
\newblock Academic Press, 1997.

\bibitem{GelfandKolmogorovYaglom:1956:OnTheGeneralDefinitionOfTheAmountOfInfor%
mation}
I.~M. Gelfand, N.~A. Kolmogorov, and A.~M. Yaglom.
\newblock On the general definition of the amount of information.
\newblock {\em Dokl. Akad. Nauk USSR}, 111(4):745--748, 1956.
\newblock (In Russian).

\bibitem{Wiener:1948:Cybernetics}
N.~Wiener.
\newblock {\em Cybernetics}.
\newblock Wiley, New York, 1948.

\bibitem{Kallianpur:1960:OnTheAmountOfInformationContainedInASingmaField}
G.~Kallianpur.
\newblock On the amount of information contained in a $\sigma$-field.
\newblock In I.~Olkin and S.~G. Ghurye, editors, {\em Essays in Honor of Harold
  Hotelling}, pages 265--273. Stanford Univ. Press, Stanford, 1960.

\bibitem{Pinsker:1960:InformationAndInformationStability}
M.~S. Pinsker.
\newblock {\em Information and Information Stability of Random Variables and
  Process}.
\newblock Holden-Day, San Francisco, CA, 1960.
\newblock (English ed., 1964, translated and edited by Amiel Feinstein).

\bibitem{KullbackLeibler:1951:OnInformationAndSufficiency}
S.~Kullback and R.~A. Leibler.
\newblock On information and sufficiency.
\newblock {\em Ann. Math. Stat.}, 22:79--86, 1951.

\bibitem{Ochs:1976:BasicPropertiesOfTheGeneralizedBoltzmann-Gibbs-ShannonEntro%
py}
W.~Ochs.
\newblock Basic properties of the generalized {B}oltzmann-{G}ibbs-{S}hannon
  entropy.
\newblock {\em Reports on Mathematical Physics}, 9:135--155, 1976.

\bibitem{Masani:1992:TheMeasureTheoreticAspectsOfEntropy_Part_1}
P.~R. Masani.
\newblock The measure-theoretic aspects of entropy, {P}art 1.
\newblock {\em Journal of Computational and Applied Mathematics}, 40:215--232,
  1992.

\bibitem{Masani:1992:TheMeasureTheoreticAspectsOfEntropy_Part_2}
P.~R. Masani.
\newblock The measure-theoretic aspects of entropy, {P}art 2.
\newblock {\em Journal of Computational and Applied Mathematics}, 44:245--260,
  1992.

\bibitem{Guiasu:1977:InformationTheoryWithApplications}
Silviu Guia{\c s}u.
\newblock {\em Information Theory with Applications}.
\newblock McGraw-Hill, Great Britain, 1977.

\bibitem{Gray:1990:EntropyAndInformationTheory}
Robert~M. Gray.
\newblock {\em Entropy and Information Theory}.
\newblock Springer-Verlag, New York, 1990.

\bibitem{Smith:2001:SomeObservationsOnTheConceptsOfInformationTheoreticEntropy}
Jonathan D.~H. Smith.
\newblock Some observations on the concepts of information theoretic entropy
  and randomness.
\newblock {\em Entropy}, 3:1--11, 2001.

\bibitem{Shannon:1948:MathematicalTheoryOfCommunication_BellLabs}
C.~E. Shannon.
\newblock A mathematical theory of communication.
\newblock {\em Bell System Technical Journal}, 27:379, 1948.

\bibitem{Jaynes:1968:PriorProbabilities}
E.~T. Jaynes.
\newblock Prior probabilities.
\newblock {\em IEEE Transactions on Systems Science and Cybernetics},
  sec-4(3):227--241, 1968.

\bibitem{ZellnerHighfield:1988:CalculationOfMaximumEntropyDistributions}
Arnold Zellner and Richard~A. Highfield.
\newblock Calculation of maximum entropy distributions and approximation of
  marginalposterior distributions.
\newblock {\em Journal of Econometrics}, 37:195--209, 1988.

\bibitem{LazoRathie:1978:OnTheEntropyOfContinuousProbabilityDistributions}
Aida C. G.~Verdugo Lazo and Pushpa~N. Rathie.
\newblock On the entropy of continuous probability distributions.
\newblock {\em IEEE Transactions on Information Theory}, IT-24(1):120--122,
  1978.

\bibitem{Ryu:1993:MaximumEntropyEstimationOfDensityAndRegressionFunction}
Hang~K. Ryu.
\newblock Maximum entropy estimation of density and regression functions.
\newblock {\em Journal of Econometrics}, 56:397--440, 1993.

\bibitem{Athreya:1994:EntropyMaximization}
K.~B. Athreya.
\newblock Entropy maximization.
\newblock IMA Preprint Series~{\tt 1231}, Institute for Mathematics and its
  Applications, University of Minnesota, Minneapolis, 1994.

\bibitem{Kantorovitz:2003:IntroductionToModernAnalysis}
Shmuel Kantorovitz.
\newblock {\em Introduction to Modern Analysis}.
\newblock Oxford, New York, 2003.

\bibitem{Csiszar:1969:OnGeneralizedEntropy}
Imre Csisz\'{a}r.
\newblock On generalized entropy.
\newblock {\em Studia Sci. Math. Hungar.}, 4:401--419, 1969.

\bibitem{Rosenblatt-Roth:1964:TheConceptOfEntropyInProbabilityTheory}
M.~Rosenblatt-Roth.
\newblock The concept of entropy in probability theory and its applications in
  the theory of information transmission through communication channels.
\newblock {\em Theory Probab. Appl.}, 9(2):212--235, 1964.

\bibitem{FerriMartinezPlastino:2005:TheRoleOfConstraintsInTsallisNonextensiveT%
reatmentRevisited}
G.~L. Ferri, S.~Mart{\'\i}nez, and A.~Plastino.
\newblock The role of constraints in tsallis' nonextensive treatmentrevisited.
\newblock {\em Physica A}, 347:205--220, 2005.

\bibitem{AbeBagci:2005:NecessityOfqExpectation}
Sumiyoshi Abe and G.~B. Bagci.
\newblock Necessity of $q$-expectation value in nonextensive statistical
  mechanics.
\newblock {\em Physical Review E}, 71:016139, 2005.

\bibitem{WadaScarfone:2005:ConnectionsBetweenTsallisFormalismEtc}
T.~Wada and A.~M. Scarfone.
\newblock Connections between {T}sallis' formalism employing the standard
  linear average energy and ones employing the normalized $q$-average enery.
\newblock {\em Physics Letters A}, 335:351--362, 2005.

\bibitem{TsallisMendesPlastino:1998:TheRoleOfConstraints}
Constantino Tsallis, Renio~S. Mendes, and A.~R. Plastino.
\newblock The role of constraints within generalized nonextensive statistics.
\newblock {\em Physica A}, 261:534--554, 1998.

\bibitem{Renyi:1960:SomeFundamentalQuestionsOfInformationTheory}
Alfred R{\'{e}}nyi.
\newblock Some fundamental questions of information theory.
\newblock {\em MTA III. Oszt. K{\"{o}}zl.}, 10:251--282, 1960.
\newblock (reprinted in~\cite{Turan:1976:SelectedPapersOfAlfredRenyi}, pp.
  526-552).

\bibitem{Renyi:1959:OnTheDimensionAndEntropyOfProbabilityDistributions}
Alfred R{\'{e}}nyi.
\newblock On the dimension and entropy of probability distributions.
\newblock {\em Acta Math. Acad. Sci. Hung.}, 10:193--215, 1959.
\newblock (reprinted in~\cite{Turan:1976:SelectedPapersOfAlfredRenyi}, pp.
  320-342).

\bibitem{Rudin:1966:RealAndComplexAnalysis}
Walter Rudin.
\newblock {\em Real and Complex Analysis}.
\newblock McGraw-Hill, 1964.
\newblock (International edition, 1987).

\bibitem{Turan:1976:SelectedPapersOfAlfredRenyi}
P{\'{a}}l Tur{\'{a}}n, editor.
\newblock {\em Selected Papers of {A}lfr{\'{e}}d {R}\'{e}nyi}.
\newblock Akademia Kiado, Budapest, 1976.

\end{thebibliography}

\end{document}